\documentclass{optica-article}

\journal{opticajournal} 

\articletype{Research Article}

\usepackage{lineno}

\usepackage{threeparttable}

\begin{document}

\newcommand{\um}{\textmu m}
\newcommand{\uA}{\textmu A}
\newcommand{\ohm}{$\Omega$}

\title{An SNSPD-based detector system for NASA's Deep Space Optical Communications project}

\author{Emma~E.~Wollman,\authormark{1,*} Jason~P.~Allmaras,\authormark{1} Andrew~D.~Beyer,\authormark{1} Boris~Korzh,\authormark{1} Marc~C.~Runyan,\authormark{1,2}  Lautaro~Narv\'aez,\authormark{3} William~H.~Farr,\authormark{1} Francesco~Marsili,\authormark{1} Ryan~M.~Briggs,\authormark{1} Gregory~J.~Miles,\authormark{1} and Matthew~D.~Shaw\authormark{1}}

\address{\authormark{1} Jet Propulsion Laboratory, California Institute of Technology, 4800 Oak Grove Dr, Pasadena, CA 91109, USA\\
	\authormark{2} Current affiliation AWS Center for Quantum Computing, Pasadena, CA 91125, USA\\
	\authormark{3} Division of Physics, Mathematics and Astronomy, California Institute of Technology, Pasadena, California 91125, USA\\}

\email{\authormark{*} emma.e.wollman@jpl.nasa.gov} 


\begin{abstract*}
We report on a free-space-coupled superconducting nanowire single-photon detector array developed for NASA's Deep Space Optical Communications project (DSOC). The array serves as the downlink detector for DSOC's primary ground receiver terminal located at Palomar Observatory's 200-inch Hale Telescope. The 64-pixel WSi array comprises four quadrants of 16~co-wound pixels covering a 320~\um\ diameter active area and embedded in an optical stack. The detector system also includes cryogenic optics for filtering and focusing the downlink signal and electronics for biasing the array and amplifying the output pulses. The detector system exhibits a peak system detection efficiency of 76\% at 1550~nm, a background-limited false count rate as low as 3.7~kcps across the array, timing jitter less than 120~ps FWHM, and a maximum count rate of $\sim$ 1~Gcps.\\ \\
© 2024. California Institute of Technology. Government sponsorship acknowledged.
\end{abstract*}

\section{Introduction}
Future space missions are expected to produce increasingly large volumes of science data over the upcoming decades. NASA is therefore exploring the possibility of complementing traditional RF communications with optical communications, which have the potential to increase data rates by a factor of 10 to 100 at Mars distances \cite{Biswas2010}. In 2013, NASA's Lunar Laser Communications Demonstration (LLCD) successfully demonstrated data rates up to 622~Mbps from lunar distances, at approximately 400,000~km \cite{Boroson2014}. The Deep Space Optical Communications (DSOC)  project is the first demonstration of optical communication from interplanetary distances \cite{Biswas2024}. At these distances, the optical signals that reach Earth are in the photon-starved regime, and the optimal method of encoding data is through the arrival time of photon pulses (pulse-position modulation, or PPM). Optical communication links therefore require ground-based receivers capable of counting single photons with high efficiency and high timing resolution. The detectors must also have a large dynamic range to enable higher data rates when the spacecraft is closer to Earth.

Over the past decade, superconducting nanowire single-photon detectors (SNSPDs) have become the detector of choice for time-correlated single-photon counting at telecom wavelengths. SNSPDs are ideal for optical communication due to their high detection efficiency (above 95\% \cite{Reddy2020}), ultra-high timing resolution (down to 3~ps \cite{Korzh2020}), and high maximum count rates (up to 1.5~Gcps \cite{Craiciu2023, Resta2023}). In addition, SNSPDs with active areas up to 1~mm$^2$ \cite{Luskin2023, Xu2023} and array formats on the scale of hundreds of kilopixels \cite{Oripov2023} have now been demonstrated.

In this paper, we describe an SNSPD-based detector system developed for NASA’s DSOC technology demonstration project. The DSOC project consists of a ground uplink terminal located at JPL's Table Mountain Facility (TMF) \cite{Srinivasan2023GLT}, a ground receiver terminal located at Caltech's Palomar Observatory \cite{Srinivasan2023GLR}, and a flight transceiver on board NASA's Psyche spacecraft. The goal of the DSOC project is to demonstrate downlink data rates ranging from 267~Mbps to 57~kbps as the Psyche spacecraft travels over a range of distances from 0.1~AU to 2.6~AU (Mars maximum range). The SNSPD array is used in the ground receiver system to convert received photons into electrical signals. A second detector system based on the same design has also been fielded at TMF as part of the Optical to Orion (O2O) project and is currently used as part of a supplementary ground receiver station for DSOC.

\section{System design}
\subsection{Array design and fabrication}
\begin{figure}[ht!]
	\centering\includegraphics[width=5.25in]{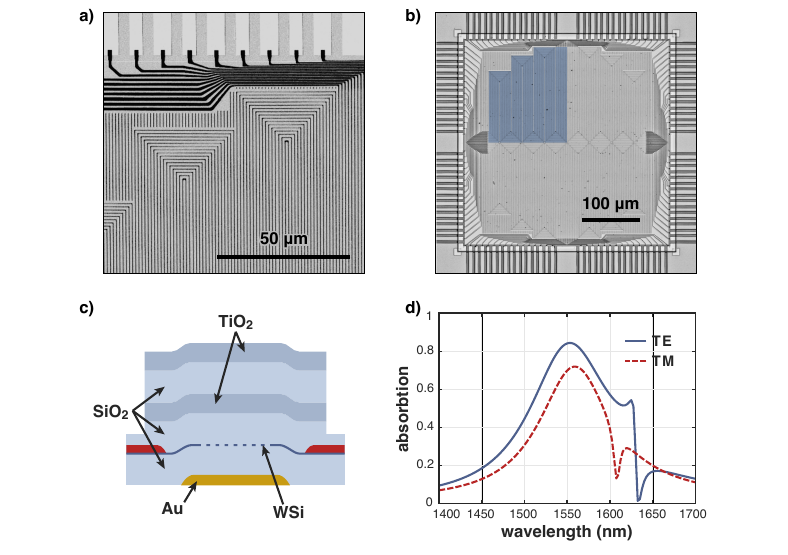}
	\caption{(a) Scanning electron microscope image of a section of one quadrant of the array. In the SEM image, the nanowire layer appears black. ``Dummy'' wire structures surround the active wires to improve fabrication uniformity. The 16 pixels of each quadrant are co-wound to minimize variations in illumination between pixels. Over the majority of the array active area, the wires form a diffraction grating oriented along the y-axis, but in regions at the edges and center of the active area where the co-wound wires bend, a horizontal grating is formed. (b) Optical micrograph of the array active area. The leads around the edge of the image form 50~\ohm\ matched CPW structures to route the nanowire signals to wire bond pads (not shown). Blue shading shows the path of the 16 co-wound wires in the upper-left quadrant. (c) Diagram of the optical stack cross-section. Layers are not to scale. The fabricated stack consists of the following layers from top (illumination side) to bottom: 166~nm TiO$_2$, 269~nm SiO$_2$, 167~nm TiO$_2$, 270~nm SiO$_2$, 4.8~nm WSi, 243~nm SiO$_2$, 2.5~nm Ti, and 80~nm Au. (d) Absorption into the nanowire grating, modeled by RCWA for light at normal incidence. TE-polarized light is polarized parallel to the nanowire grating, and TM-polarized light is perpendicular.}
	\label{fig:design}
\end{figure}

In order to meet the requirements of the DSOC project, the SNSPD array design must allow for efficient coupling to Palomar Observatory's 200-inch Hale telescope while maintaining the ability to count at high rates. There is a fundamental tradeoff between these two goals, because increasing the length of a nanowire in order to cover a larger area increases its kinetic inductance and thus slows down its recovery time. The SNSPD array design is based around a 320~\um\ diameter active area, which allows for a 50~\textmu rad field of view on the sky with a numerical aperture of 0.37 at the detector. In order to achieve high count rates and to allow for multiple detections per PPM slot, the active area is distributed across 64 SNSPD nanowires. The active area is divided into four quadrants to allow for centroiding of the incoming beam. Within each quadrant, the 16 nanowires are co-wound to fill the active area. The use of co-wound pixels minimizes the photon flux variation between pixels in order to take full advantage of the maximum count rates of all pixels in a quadrant. The fill factor of the array was optimized to maximize the optical absorption while minimizing the kinetic inductance of the wires and crosstalk between adjacent pixels. For a co-wound WSi nanowire array with 160~nm wide and 4.8~nm thick nanowires, we found that a 1200~nm pitch optimized the tradeoff between efficiency and active area without introducing crosstalk between adjacent nanowires (Fig.~\ref{fig:design}a). Due to the co-wound design and large pitch, the active area is not perfectly circular. Fig.~\ref{fig:design}b shows an optical micrograph of the array with one quadrant shaded to highlight the shape of the active area.

In order to enhance optical absorption into the nanowire layer despite its relatively low 13.3\% fill factor, we used an optical cavity consisting of an anti-reflective (AR) coating and a gold mirror to maximize coupling efficiency at a wavelength of 1550~nm (Fig.~\ref{fig:design}c). Rigorous coupled wave analysis (RCWA) \cite{Moharam1981,Zhang2010,Zhang2012} was used to calculate the expected absorbance in the device and determine the necessary thicknesses of the dielectric layers to maximize optical coupling. As the stack was fabricated, the design thicknesses of the remaining layers were adjusted to correct for any thickness errors in the previously deposited layers of the optical stack. The dielectric layer thicknesses were measured by ellipsometry and reflectometry on Si witness chips while the thicknesses of the WSi, Ti, and Au were estimated from deposition time. The modeled absorption into the final optical stack is shown in Fig.~\ref{fig:design}d. 

To begin fabrication, we deposited the Au mirror layer for lift-off by electron beam evaporation on a 100-mm Si wafer. SiO$_2$ was sputtered on a Ti adhesion layer under RF bias to form a quarter-wavelength dielectric layer on the mirror. WSi was sputtered from a compound target leading to a film with a resistivity of $1.8\times10^{-6}$~\ohm$\cdot$m at 300~K. Electrical contacts were patterned by optical lithography to form 50~\ohm\ matched coplanar waveguides (CPWs) to route the nanowire signals to the device bonding pads. After optical lithography, the nanowires were patterned using electron beam lithography on a negative tone (ma-N 2401) electron beam resist and by etching in CHF$_3$/O$_2$ plasma. Additional inactive nanowire features were patterned around the perimeter of the active area to minimize proximity effects during exposure of the electron-beam resist and ultimately achieve a more uniform wire width across the array. To further ensure uniform features, a spatially-varying electron-beam dose was applied based on modeled proximity effects resulting from the electron-beam parameters, the composition of the underlying layers, and the local pattern geometry. After nanowire patterning, the dielectric layers were sputtered to build the anti-reflection layers. 

After several iterations of design optimization and process development, a prototype array was fabricated that appeared to meet all the requirements of the DSOC project. At this point, three wafers of detectors were fabricated using the same array geometry and optical stack, and three dies from each of the wafers were packaged and screened on a testbed with prototype optics and electronics. Of the nine screened arrays, seven met the requirements of the DSOC project, and four reached saturated efficiency in all 64 channels. Of these four, we chose a primary and a backup array for the DSOC project and for the O2O project. The results reported here are either from the prototype array, the primary DSOC array, or the primary O2O array. The prototype array differs from the later devices in its packaging and in the layout of its leads. There is also some variation in the thickness of the WSi layer between devices from different wafers that leads to small differences in performance. The properties of all screened detectors are listed in the Supplemental Material.

\subsection{Detector assembly design}
In the DSOC project structure, the SNSPD array is one component of the larger ground detector assembly (GDA). In turn, the GDA is part of the ground laser receiver system (GLR), which is located in the Hale Telescope's Coudé Room. The detector assembly contains a cryostat, cryogenic optics for filtering and focusing light on the detector, and cryogenic and room-temperature electronics for routing the array's electrical signals and amplifying its output. The GDA also contains an optical test stimulus source (OTSS), which is able to produce both modulated and CW light at calibrated power levels. The OTSS is used for detector and system calibration and characterization. The detector assembly interfaces with the ground laser receiver optical assembly (GLROA), which is responsible for routing light from both the telescope and OTSS to the detector. The GLROA contains a near-IR camera for acquisition of the downlink signal, narrowband filters for rejection of sky background, and a variable zoom system to optimize the detector field of view. The detector assembly also interfaces with the ground signal processing assembly (GSPA), which time stamps the array pulses using a 64-channel time-to-digital converter (TDC), converts the time tags to PPM symbols, and decodes the PPM data in real time. All assemblies have their own internal monitor and control (M\&C) systems that interface with the global GLR M\&C software. Further information on the GLR system design can be found in Srinivasan et al. \cite{Srinivasan2023GLR}.

\begin{figure}[ht]
	\centering\includegraphics[width=5.25in]{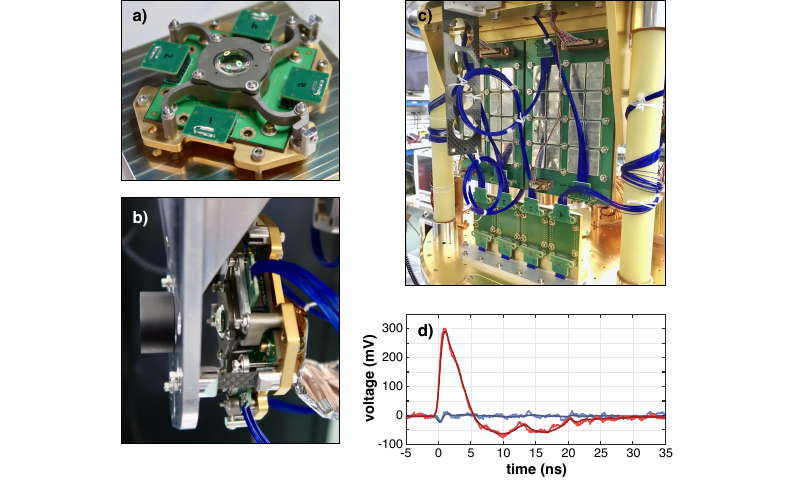}
	\caption{(a) Packaged array with cryogenic lens in a titanium flexure mount. Spring-loaded screws on two arms of the flexure mount allow for X/Y adjustment, and shims inserted between the lens holder and the flexure mount adjust focus. Any residual misalignment can be compensated for by the GLROA. (b) Array mounted in the cryostat with lens and filter. The bracket at left is attached to the 3~K stage, and the detector plate is thermalized to the 1~K stage using the copper foil strap on the right. Carbon-fiber standoffs provide thermal isolation. One of the short-pass filters is mounted in a lens tube on the 3~K bracket. (c) Cryogenic electronics, including HLCD cables, 3~K thermalization boards (bottom), and two 16-channel cryogenic amplifier boards (center). Two more amplifier boards are mounted on the backside of the 40~K bracket. (d) SNSPD pulse at the output of the amplifier chain. The light red trace shows a single pulse, and the dark red trace represents the time-averaged pulse shape. Likewise, the light blue trace shows the signal on an adjacent channel, and the dark blue trace shows the time-averaged electrical crosstalk. The crosstalk is not large enough to cause spurious counts, but it can distort the pulse shape if adjacent channels trigger at the same time, which can lead to increased jitter.}
	\label{fig:electrical}
\end{figure}

The detector assembly cryostat consists of a modified FormFactor Model 106 cryostat with a Cryomech PT410 pulse tube and a Chase GL-4 4He sorption refrigerator. The 1~K stage of the refrigerator cools the detector to a temperature of 960~mK. Light is free-space-coupled onto the SNSPD array through three cryogenic filter windows mounted in the 40~K radiation shield, the 3~K radiation shield, and a 3~K bracket. Filtering of 300~K blackbody emission from room temperature optics is necessary to reduce the array’s dark count rate, because the nanowires are sensitive to wavelengths out to $\sim$ 4~\um. The custom filters from Andover Corporation consist of reflective short-pass coatings on half-inch substrates of BK7, which is absorptive at wavelengths above 2.3~\um. The filter coatings in the radiation shields have a cutoff of 1.9~\um, and the filter at the 3~K bracket has a cutoff of 1.6~\um. Further details about the filters can be found in Mueller et al. \cite{Mueller2021}. The window through the 300~K stage of the fridge is 2-inch diameter, 12~mm thick, AR-coated BK7 (Thorlabs WG12012-C). In the DSOC system, a cryogenic lens (Thorlabs AL1815-C) is mounted to the detector plate to provide a large enough NA for coupling light from the telescope’s 5~m aperture onto the detector’s 320~\um\ area. Figure~\ref{fig:electrical}a shows the cryogenic lens integrated with the detector, and Figure~\ref{fig:electrical}b shows the detector/lens assembly installed in the cryostat. The OCTL telescope used for O2O has a 1~m aperture, allowing for the final focusing lens to be located outside of the cryostat.

Each pixel of the array is biased and read out individually. Routing of RF signals inside the cryostat is performed using high-flexibility micro-coax cables with custom high-density connectors (Samtec HLCD and LSHM series). Metal-core PCBs at 3~K and at the sorption refrigerator's film-burner stage are used to thermalize the cables. At 40~K, four 16-channel amplifier boards provide two stages of amplification using PHEMT (Avago Technologies ATF 35143) and SiGe (RFMD SGL0622Z) amplifiers. The cryogenic electronics are shown in Fig.~\ref{fig:electrical}c. Following amplifiers (Minicircuits RAM 8A+) at room temperature are used to increase the final pulse amplitude. The input of the PHEMT stage is DC-coupled with a 50-\ohm\ termination to avoid the re-biasing challenges associated with AC-coupled amplifiers operating at high count rates \cite{Kerman2013}. The cryogenic amplifiers high-pass the SNSPD pulses, shortening the pulse width from >~20~ns to $\sim$5~ns (Fig.~\ref{fig:electrical}d). Some electrical crosstalk was observed between channels, particularly nearest neighbors. The high-density cable connectors were identified as the primary contributor to the electrical crosstalk, and the connector PCBs were redesigned to improve isolation between adjacent channels. While some crosstalk remains, it is below the comparator threshold level and does not lead to extra counts. It can, however, impact the timing jitter when there are many photons per laser pulse due to a voltage shift in the SNSPD output.

The outputs of the room-temperature amplifiers are coupled into 64 SMA coaxial cables. These are connected to the GSPA's 64-channel time-to-digital converter (Dotfast Solutions TDM1600-64). The TDC triggers on SNSPD pulses using a fixed-threshold comparator front-end and outputs sorted time tags with a timing resolution of 15.625~ps and full-width half-maximum (FWHM) timing jitter below 50~ps. Time tags are streamed over PCI Express either to disk or to the receiver FPGAs at transfer speeds up to 1.5~GTag/s.

Each SNSPD channel is individually biased using an NI PXI DAC voltage source (NI PXIe-6739) and a 200~k\ohm\ resistor at room temperature. A resistive bias tee couples the bias current to the detector at the input of the cryogenic amplifier. Cable resistance between the nanowire and input of the cryogenic amplifier leads to current splitting between the device and the 50~\ohm\ input termination of the amplifier. Because the cryogenic series resistance is unknown, we report the total bias current drawn by each channel of the voltage source. The actual current in the device is lower. The testbed used for screening the arrays had different cables between the bias tee and device, so the splitting ratio was different between the screening measurements and measurements in the GDA cryostat as installed at Palomar Observatory.

\section{System performance}

\subsection{Detection efficiency}\label{sec:sde}
The system detection efficiency of the array ($SDE_a$) is defined as the fraction of photons entering the room-temperature window of the cryostat that are counted by the readout electronics. The array system detection efficiency is calculated by measuring the total count rate across the array with and without a shutter blocking the light source. The array background count rate ($BCR_a$) with the source shuttered is subtracted from the array count rate with the source unshuttered to produce the array photon count rate ($PCR_a$). $SDE_a$ is the ratio of $PCR_a$ to the rate of signal photons incident on the cryostat. It is also useful to define the pixel-level equivalents $SDE_p$, $BCR_p$, and $PCR_p$, defined such that $SDE_a = \sum{SDE_p}$. 

\begin{figure}[h]
	\centering\includegraphics[width=5.25in]{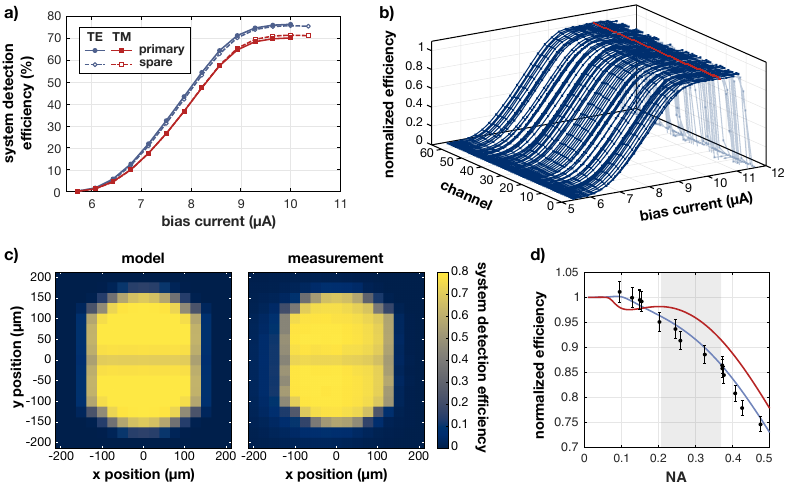}
	\caption{(a) Array system detection efficiency vs. bias current for the DSOC primary array and the DSOC spare, measured with a low-NA lens on the screening cryostat at a temperature of 900~mK. Both detectors show a saturated efficiency of $\sim$75\% for TE-polarized light and $\sim$70\% for TM-polarized light at 1550~nm. (b) $PCR_p$ vs. bias current for all pixels of the primary DSOC array, measured with a CW laser at 1550~nm in the GDA cryostat. All 64 pixels exhibit saturated internal detection efficiency. The photo-count rate at the operating current of 10.5~\uA\ is highlighted in red. (c) Model (left) and measurement (right) of the spatial dependence of the detection efficiency for a 50~\um\ spot. The regions where the wires bend produce a lower detection efficiency near $y=0$ in both the model and measurement. Differences between the two may be due to vibrations of the cryostat during measurements. (d) Array system detection efficiency vs. numerical aperture of the focused beam, normalized to the low-NA efficiency. A change in incident angle results in an effective shift of the optical cavity wavelength and a decrease in the absorption in the nanowire layer. The range of NAs used in the DSOC optical system is indicated by the shaded region.}
	\label{fig:pcr}
\end{figure}

The incident photon flux is measured by inserting a mirror in front of the cryostat to deflect the beam onto a free-space power meter. A high flux is first applied to provide a reference power, and then calibrated attenuation is applied to reduce the flux to the desired level. When calculating the $SDE$, the measured power must be adjusted for the additional loss from the extra mirror and lens used for the free-space power measurement. The total uncertainty in the efficiency measurement is $\pm 5\%$. In the screening testbed, a similar procedure is used to calibrate the incident photon flux. However, in the testbed, a sliding breadboard for the free-space optics allows for measurement of the free-space power without any additional optical components. Fig.~\ref{fig:pcr}a shows the efficiency of the primary and spare DSOC arrays measured for both TE- and TM-polarized light at $\lambda = 1550$~nm using a low-NA lens in the screening testbed. Both arrays have a maximum efficiency of 76\% for TE light and $\sim$ 70\% for TM light. The DSOC downlink signal is circularly-polarized, but is converted to linear TE polarization by a quarter-waveplate and a half-waveplate in the GLROA. The polarization of the O2O downlink signal is not controlled.

Fig.~\ref{fig:pcr}b shows the normalized $PCR_p$ as a function of bias current ($I_b$) for all 64 pixels of the primary DSOC array at a temperature of 950~mK on the GDA cryostat. An efficiency plateau, indicating saturated internal detection efficiency, is present for all pixels. During normal operations, all pixels of the array are biased mid-plateau at a current of 10.5~\uA. The array is biased in the middle of the plateau to allow for potential variations in temperature or switching current over time, although no significant variations have been observed to date.

For single-mode-fiber-coupled SNSPDs, the incoming beam has a well-defined numerical aperture and spot diameter. In contrast, the free-space-coupled DSOC array must be able to accomodate different spot sizes due to varying atmospheric conditions, and it must accept a large NA beam for efficient coupling to the large telescope area. The array's optical coupling efficiency depends on both the spot size and NA of the optical signal, so the GLROA contains an adjustable zoom system to optimize the signal on the detector for different seeing conditions. The different contributions to the array efficiency are discussed in detail in the Supplemental Material. 

Fig.~\ref{fig:pcr}c illustrates how the array layout leads to a dependence of the efficiency on the spot size and position. The map on the left shows the modeled TE efficiency as a function of position on the array for a 50~\um\ diameter spot, and the map on the right shows the corresponding measurement performed by scanning a tightly-focused beam across the active area with a fast steering mirror. The region of bends in the center of the array has a nanowire orientation perpendicular to the majority of the active area, leading to a decrease in absorption for TE-polarized light in both the model and measurement. If the spot size is too small, it will predominantly sample the bend region and lead to a lower $SDE_a$. If the spot size is too large, the efficiency will be lower due to overfilling. The optimal spot size for the array is therefore between 90 and 250~\um\ in diameter.

For the DSOC project, a variable zoom system adjusts the detector field of view from 27~\textmu rad to 50~\textmu rad (NA = 0.2 to 0.37) to accommodate different atmospheric seeing conditions. The optical cavity for the SNSPD array was optimized for normal incidence. When light comes in at an angle, the center wavelength of the cavity is effectively shifted, and the absorption into the nanowire layer is reduced. Fig.~\ref{fig:pcr}d shows the resulting decrease in absorption modeled for TE and TM polarized light as a function of numerical aperture. The NA dependence was also measured in the screening testbed for TE light using a cryogenic lens and different incoming beam diameters. The Supplemental Material includes a full description of the angular dependence. Under nominal seeing conditions, the efficiency of the DSOC array is expected to be 70\%. Lab measurements of the efficiency using a phase plate to emulate nominal seeing conditions confirm this prediction (measured values between 69 and 72\%).

\subsection{Dark count rate}
\begin{figure}[h!]
	\centering\includegraphics[width=5.25in]{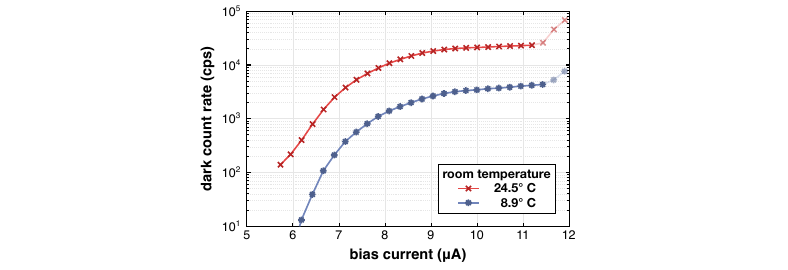}
	\caption{Dark count rate of the DSOC array vs. bias current measured in summer (red) and winter (blue). At the operational bias point of 10.5~\uA, the dark count rate ranges from 3.7~kcps to 22~kcps. Above a bias current of 11.25~\uA, channels begin to exhibit relaxation oscillations or latching.}
	\label{fig:dcr}
\end{figure}

The dark count rate of the array is dominated by room temperature blackbody radiation; with the 3~K window blanked, $BCR_a$ is on the order of 1 - 10~cps, and is likely still limited by residual stray light inside the cryostat. The dark count rate therefore depends on the detector's 300~K field of view and on the cryogenic filters used. In the GDA cryostat, the 300~K field of view is set by the cryogenic lens, which has an NA of 0.53. The detector's frequency-dependent optical efficiency (Fig.~\ref{fig:design}d) provides additional filtering. Because the Hale Telescope's Coudé room is not temperature-controlled, the dark count rate also depends on the temperature of the room. Fig.~\ref{fig:dcr} shows the array dark count rate vs. bias current measured in the summer (T = $24.5^\circ$ C) and in the winter (T = $8.9^\circ$ C). At the DSOC operating current, the dark count rate was 22~kcps in the summer and 3.7~kcps in the winter. Total nighttime on-sky background count rates are a few hundred kcps without additional filtering or 10 - 50~kcps with a 1.8~nm bandpass filter installed in the GLROA.

\subsection{Maximum count rate}
\begin{figure}[h!]
\centering\includegraphics[width=5.25in]{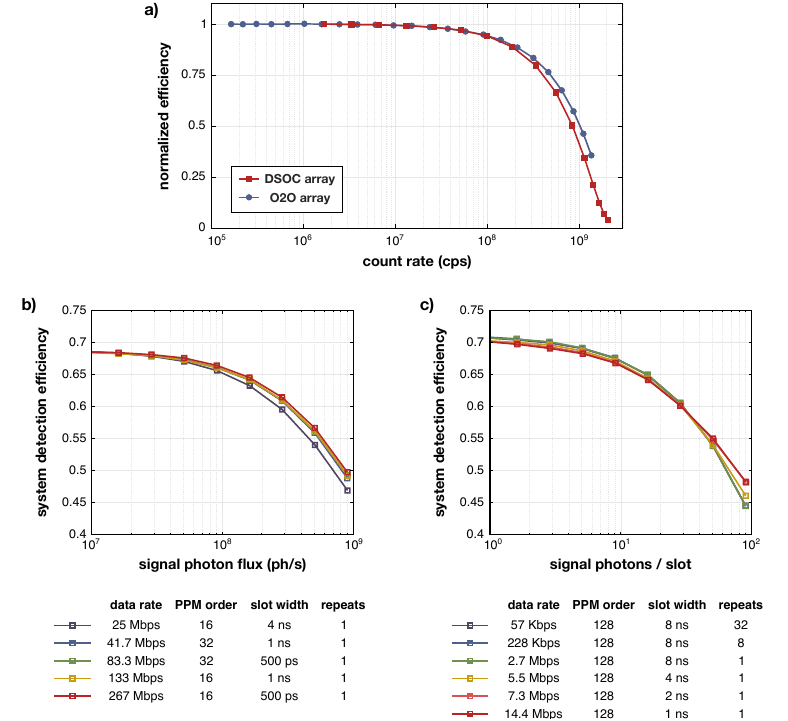}
\caption{a) Normalized $SDE_a$ vs. count rate under CW flood illumination for the DSOC array biased mid-plateau (red) and the O2O array biased just below the switching current (blue). The 3-dB saturation points are 850~Mcps and 1~Gcps, respectively. b) Measured $SDE_a$ vs. input photon flux for different PPM data formats. All data formats have symbol periods < 100~ns. c) Measured $SDE_a$ vs. number of input signal photons per symbol for different PPM data formats. All data formats have symbol periods > 100~ns. For both b) and c), the array was illuminated using a spot size and NA representative of nominal seeing conditions at Palomar.}
\label{fig:mcr}
\end{figure}

The DSOC array must be able to count at rates of several hundred Mcps in order to match the received photon flux at the highest data rates. Each of the 64 nanowires in the array has a dead time of approximately 20 ns, after which it recovers to full efficiency over a period of approximately 80 ns. Photons that are absorbed within 20 ns after a detection event in the same wire are not detected, and photons that are absorbed within 20-100 ns of a detection event in the same wire are detected with a lower probability. Fig.~\ref{fig:mcr}a shows the normalized array detection efficiency vs. count rate under CW flood illumination. The 3-dB saturation count rate is 850~Mcps when biased in the middle of the $PCR$ plateau. A higher maximum count rate can be achieved by biasing the array at a higher bias current. For example, the O2O array has a 3-dB saturation count rate above 1~Gcps when biased close to the switching current. Measurements of count rates above $\sim$2~Gcps are limited by the TDC counter's maximum count rate.

For PPM data formats with short symbol periods (< 100 ns), the blocking loss primarily depends on the rate of incoming photons, because typical inter-pulse separations are less than the detector dead time. Fig.~\ref{fig:mcr}b shows the measured system detection efficiency vs. incoming photon flux for different data formats where the symbol period is < 100~ns. The curves are similar to the measurement under CW illumination. For data formats with long symbol periods (> 100 ns), however, pulses are typically separated by more than the detector dead time, so the blocking loss primarily depends on the number of photons per pulse – the array misses photons when multiple photons in the pulse are incident on the same nanowire. Fig.~\ref{fig:mcr}c shows the system detection efficiency vs. number of incoming signal photons per slot for different data formats where the symbol period is > 100~ns. The measurements for the different data rates follow the same curve despite representing very different count rates, indicating that the blocking loss is dominated by photon number rather than photon rate.

\subsection{System jitter}
\begin{figure}[h!]
	\centering\includegraphics[width=5.25in]{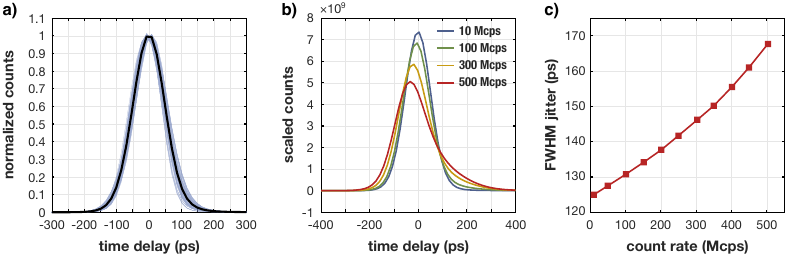}
	\caption{a) Scaled jitter histograms for individual channels (blue) and for the whole array (black) measured with a 20~MHz pulsed laser. b) Array jitter histograms at different count rates measured with a 500~MHz pulsed laser. c) FWHM jitter vs. count rate, measured with the 500~MHz laser.}
	\label{fig:jitter}
\end{figure}

The jitter of the array, GDA electronics, and TDC was characterized using a 20~MHz mode-locked laser. The laser's 20~MHz electrical sync signal was used to generate a 10~MHz clock reference for the TDC using a PLL-based clock translator (AD9553). The TDC's comparator level was set at 25\% of the pulse height for each channel. The optimal trigger level is 45\%, but using a lower threshold helps with the effects of temporal walk, as discussed below. One set of time tags was saved and analyzed to produce a time delay calibration for the TDC. Another set of time tags was then saved with the calibration loaded in the TDC. Fig.~\ref{fig:jitter}a shows jitter histograms for each channel of the array, obtained from the second time tag acquisition by binning the time tags modulo the laser period. The mean FWHM jitter of the individual channels is 118~ps. A histogram of counts from all channels of the array is plotted in red in Fig.~\ref{fig:jitter}a, and its FWHM is also 118~ps. The clock derivation from the laser sync is estimated to contribute $\sim$32~ps of jitter, and the TDC jitter is approximately 50~ps. The jitter is expected to be dominated by measurement noise. A prototype version of the DSOC array was measured to have single-channel jitter of 40~ps FWHM using a low-noise cryogenic amplifier and a fast oscilloscope.

At high count rates, the array jitter increases. The degredation in jitter is due to variations in the pulse shape that lead to timing offsets when combined with the TDC's fixed-threshold comparator \cite{Mueller2023}. These variations have several potential sources: smaller pulse amplitudes occur when the bias current has not fully returned to the nanowire following the previous detection; shifted pulses can occur when voltage ripples from a previous event's pulse interfere with the current pulse; and distortions in the rising edge of the pulse can occur due to electrical cross-talk from a neighboring pixel that detects a photon at the same time. All these effects are more severe when the average time between events is shorter or if more photons are detected in the array at the same time. As part of the technology development phase of the project, the readout electronics were optimized to minimize the count-rate-dependent jitter by speeding up the cryogenic amplifiers so that the undershoot in the electrical pulse (Fig.~\ref{fig:electrical}d) occurs within the detector's recovery time and by redesigning the micro-coax connectors with additional grounding between signal lines to minimize cross-talk. Fig.~\ref{fig:jitter}b shows histograms measured using a 500~MHz pulsed laser at different count rates. As the count rate increases, the histogram width increases, and it develops a longer tail at higher delay values. Fig.~\ref{fig:jitter}c shows the FWHM jitter vs. count rate measured with the 500~MHz laser. The FWHM jitter increases by about $35\%$ from its low-count-rate limit at a count rate of 500~Mcps, and the full width at 1\% increases even more severely. We investigated the possibility of correcting for the time walk contribution to the jitter, as described in \cite{Mueller2023}, but did not find a significant improvement in decoding performance. The GSPA includes a matching filter for jitter compensation, which, in conjunction with the error correction built in to the Consultative Committee for Space Data Systems (CCSDS) code standard \cite{CCSDS-coding}, is able to handle the increase in jitter at high count rates.

\subsection{Crosstalk}
In a photon-counting array, crosstalk occurs when detection events on one channel cause detection events on another channel. Crosstalk in SNSPD arrays can be due to either electrical or thermal coupling between the channels. For example, in early prototypes of the DSOC array design, the nanowires were spaced too densely, and the heat produced by a detection event on one channel was able to propagate to neighboring channels and trigger an event several nanoseconds later. In the current DSOC array, electrical coupling produces small negative pulses on neighboring channels when a channel detects a photon (Fig.~\ref{fig:electrical}d), but these pulses are below the comparator threshold level and therefore do not cause crosstalk events. In general, crosstalk mechanisms have a characteristic time scale of correlations between channels. In the absence of crosstalk, events on different channels will be completely uncorrelated. 

\begin{figure}[h!]
\centering\includegraphics{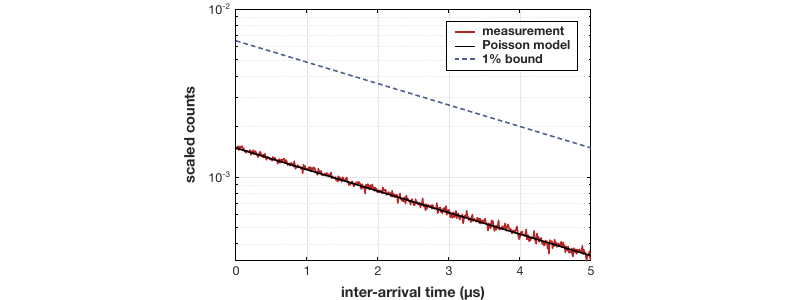}
\caption{Inter-channel correlation measurement used to check for crosstalk. The probability of detecting an event on one channel following an event on the adjacent channel is plotted as a function of time between events. Red solid line: measured values, binned by 10~ns. Black line: un-correlated Poisson model prediction. Blue dashed line: upper limit of correlations in a 10~ns window with $1\%$ crosstalk, calculated from the measured average count rates. The measured values follow the Poisson-model prediction with no indication of crosstalk.}
\label{fig:xtalk}
\end{figure}

To look for timing correlations between channels, we collected over 100 million time tags across the whole array under CW illumination and looked at the interarrival times between events on adjacent channels. The simplest way to analyze the interarrival times is first to isolate tags from the two channels of interest (e.g. Ch1 and Ch2), and then look for pairs of consecutive events when Ch1 clicked and then Ch2 clicked. Correlation in the time separation of these pairs is an indication of crosstalk on Ch2 due to events on Ch1. The same analysis is repeated for crosstalk on Ch1 due to events on Ch2. Fig.~\ref{fig:xtalk} shows a representative interarrival time histogram for two adjacent channels of the DSOC array. The histogram follows an exponential decay as expected for Poisson-distributed events without any evidence of deviations due to crosstalk. The measurement sensitivity was better than the DSOC project's requirement of $<1\%$, but no crosstalk was observed. All adjacent channel pairs were analyzed, and none showed signs of crosstalk. Correlations between Ch8 and all other channels in the same quadrant were also analyzed to look for any crosstalk between non-adjacent channels with similar findings.

\section{Deployment}
The detector system was first assembled and tested at the Jet Propulsion Laboratory. The system was then delivered to Palomar Observatory in July 2021 and installed in the Coud\'e room of the Hale Telescope. The cryostat operated continuously from August 2021 until November 2022 and from May 2023 until July 2024, with a few outages due to planned facility maintenance  to unplanned loss of power or cooling water. During the extended pre-launch operations period, the GLR underwent various on-sky and off-sky tests in order to verify its ability to acquire and track on-sky sources in different seeing conditions, to refine its operating procedures and control software, and to improve its robustness to unplanned utility outages.

\begin{figure}[h!]
	\centering\includegraphics{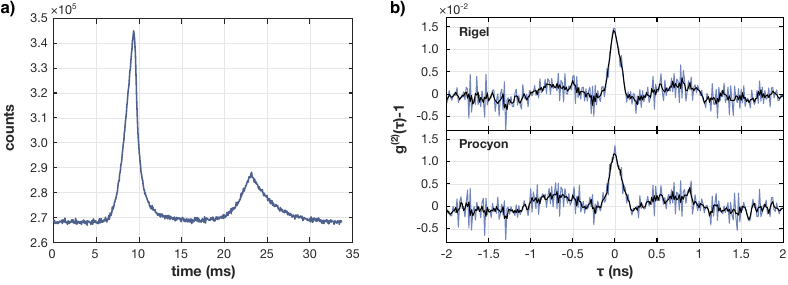}
	\caption{Astronomical observations using the DSOC instrument. a) Light curve of the Crab Pulsar measured using the DSOC SNSPD array on October 21, 2021. The best-fit period from 20~minutes of measurements is 33.7916~ms. b) g$^{(2)}$  measurements of two bright stars using the DSOC SNSPD array. Correlations were measured between two halves of the array. Blue traces show the g$^{(2)}$  histograms using the TDC resolution of 15.625~ps, and black traces show filtered histograms. Counts for Rigel were collected for a total of 123~seconds at a total count rate of 43.9~Mcps. Counts for Procyon were collected for a total of 40~seconds at a count rate of 80.5~Mcps.}
	\label{fig:crab}
\end{figure}

On October 21, 2021, the GLR was used to measure a light curve of the Crab Pulsar. A 1450~nm long-pass filter was used to reduce background. The Crab Nebula was too faint to allow for centroiding on the detector, so the telescope was blind-pointed to the nebula coordinates after centering on a nearby star. 20 minutes of time tags were collected and processed. The time tags were folded by the best-fit period of 33.7916~ms and binned by 50~\textmu s. Due to pointing uncertainty, the pulsar only contributed to counts on half of the array, so counts from the other two quadrants were discarded. Fig.~\ref{fig:crab}a shows the resulting light curve.

To demonstrate the large collection area, high maximum count rate, and high timing resolution of the GLR, the second-order autocorrelation function was measured for several bright stars to demonstrate thermal photon bunching. Observations were conducted using a 1550~nm bandpass filter with a 1.8~nm nominal bandwidth. Instead of using a beam splitter and two detectors as in the original Hanbury Brown and Twiss measurements \cite{HanburyBrown:1956}, we measured correlations between two halves of the array. Fig.~\ref{fig:crab}b shows g$^(2)$ measurements for Rigel and Procyon, with a clear increase in correlations at $\tau=0$. This demonstration suggests that SNSPDs would make good candidates for infusion into astrometrical intensity interferometers, which currently use SPAD or PMT detectors \cite{Guerin2018, Abeysekara2020, Zampieri2021}.

DSOC operations started in November 2023, and the GLR was successfully able to decode transmitted data at rates up to the maximum supported data rate of 267~Mbps, achieving this data rate at distances up to 0.37~AU (equivalent to the minimum Earth-Mars distance). At a distance of 1.5~AU (equivalent to the average Earth-Mars distance), the GLR was able to decode at a data rate of 25~Mbps. At a distance of 2.68~AU (equivalent to Mars farthest range), the GLR was able to decode at 8.33~Mbps. The similar O2O detector system was also used as part of a back-up ground receiver at the 1~m OCTL telescope to decode at data rates up to 61.25~Mbps at a range of 0.13~AU. During the first year of operations, the flight laser was limited to half of its maximum power and lower PPM orders as a risk reduction measure. Operations are scheduled to continue into 2025 with the possibility of increasing laser power and using higher PPM order formats.

\section{Conclusions and future work}
We have reported on an SNSPD-based detector system for the DSOC Ground Laser Receiver. The 64-channel detector array has a detection efficiency of 70\% under nominal seeing conditions, total dark counts as low as 3.7~kcps, a maximum count rate of $\sim$ 1~Gcps at the 3-dB saturation point, and jitter of 118~ps FWHM. As part of the larger DSOC Ground Laser Receiver system, the detector assembly enabled links at data rates up to 267~Mbps. For the next generation of deep-space optical communication ground receivers, even larger and faster SNSPD arrays will be needed, and readout of future systems remains a challenge. The current approach of direct readout for each channel will become impractical for arrays much larger than 100 pixels due to cryogenic heat loads and readout complexity. Different approaches to both increase the speed of each channel and to reduce the number of readout channels through cryogenic multiplexing will likely be necessary. While challenges remain in the continued scaling of SNSPD arrays, we have demonstrated that the technology has matured to the point of producing large-scale systems that meet the demands of deep-space optical communication applications.

\section*{Funding}
The research was carried out at the Jet Propulsion Laboratory, California Institute of Technology, under a contract with the National Aeronautics and Space Administration (80NM0018D0004).

\section*{Acknowledgments}
The authors would like to thank the DSOC project for their support, including project manager Bill Klipstein, project technologist Abi Biswas, ground manager Meera Srinivasan, and additional members of the Ground Laser Receiver team: Erik Alerstam, Ryan Rogalin, Nathaniel Richards, Angel Velasco, Seán Meenehan, Roger O'Brient, Carlos Esproles, Vachik Garkanian, Huy Nguyen, and Sabino Piazolla. The cryogenic amplifier development was made possible by a collaboration with the Spiropulu group / INQNET at Caltech. We thank our partners from Caltech Optical Observatories at Palomar Observatory for their support and contributions to the DSOC project. The authors would also like to thank Thomas Lehner of Dotfast Solutions, Andrew Rathbone and Brian Stoddard of Cryomech, Inc., and Vikas Anant of Photon Spot, Inc. for the extra time they contributed in helping us to adapt their products for our unique needs. The authors acknowledge helpful discussion and advice from collaborators at NIST and MIT Lincoln Labs. The authors would like to credit Jeffrey Stern (1962 – 2013) with performing foundational work on SNSPD development at JPL. The research was carried out at the Jet Propulsion Laboratory, California Institute of Technology, under a contract with the National Aeronautics and Space Administration. This work was supported in part by a NASA Space Technology Research Fellowship.


\bibliography{dsoc}

\end{document}